\documentclass{article}
\usepackage{spconf,amsmath,graphicx}

\usepackage[utf8]{inputenc} 
\usepackage[T1]{fontenc}    
\usepackage{url}            
\usepackage{booktabs}       
\usepackage{amsfonts}
\usepackage{amsmath,amssymb,graphicx}
\usepackage{enumerate,amsthm}
\usepackage{bbm}
\usepackage{dsfont}
\usepackage{algpseudocode}
\usepackage{algorithmicx}
\usepackage{algorithm}
\usepackage{comment}
\usepackage{color}
\usepackage{multirow}

\usepackage{verbatim}

\newcommand{\argmin}{\operatornamewithlimits{argmin}}
\newcommand{\argmax}{\operatornamewithlimits{argmax}}

\def\x{{\mathbf x}}
\def\y{{\mathbf y}}
\def\z{{\mathbf z}}

\def\bR{{\mathbb R}}

\def\NPDF{{\mathcal N}}

\def\f0{{\mathbf 0}}

\def\prox{{\operatorname{prox}}}

\theoremstyle{definition}

\title{THE INCREMENTAL PROXIMAL METHOD: A PROBABILISTIC PERSPECTIVE}
\name{\"Omer Deniz Aky{\i}ld{\i}z$^\star$\thanks{Corresponding e-mail: \texttt{omerdeniz.akyildiz@uc3m.es}.}, V\'ictor Elvira$^\dagger$, Joaqu\'in M\'iguez$^\star$}
\address{$^\star$Dept. of Signal Theory and Communications, Universidad Carlos III de Madrid, Legan\'es, Spain. \\
$^\dagger$IMT Lille Douai \& CRIStAL (UMR CNRS 9189), Villeneuve d'Ascq, France.}
\begin{document}
%
\maketitle
\begin{abstract}
In this work, we highlight a connection between the incremental proximal method and stochastic filters. We begin by showing that the proximal operators coincide, and hence can be realized with, Bayes updates. We give the explicit form of the updates for the linear regression problem and show that there is a one-to-one correspondence between the proximal operator of the least-squares regression and the Bayes update when the prior and the likelihood are Gaussian. We then carry out this observation to a general sequential setting: We consider the incremental proximal method, which is an algorithm for large-scale optimization, and show that, for a linear-quadratic cost function, it can naturally be realized by the Kalman filter. We then discuss the implications of this idea for nonlinear optimization problems where proximal operators are in general not realizable. In such settings, we argue that the extended Kalman filter can provide a systematic way for the derivation of practical procedures.
\end{abstract}
\section{Introduction}
In signal processing and machine learning, it is often of interest to solve unconstrained optimization problems of the form
\begin{align}\label{TheCost}
\min_{\theta \in \bR^{d}} f(\theta) = \min_{\theta \in \bR^{d}} \sum_{k=1}^n f_k(\theta),
\end{align}
where $f(\theta)$ is the cost function that is built up from the additions of a large number, $n$, of terms given by the functions $f_k$. The parameter to be optimized, $\theta$, is a real vector of dimension $d$.

The setting of Eq.~\eqref{TheCost} typically arises when one has a large number of independent observations. When $n$ is large, it is not possible to use classical first and second order optimization algorithms, either because gradients are too expensive to compute or Hessian matrices are impossible to store. Stochastic optimization methods have emerged as a powerful solution to this problem. Among many methods, stochastic gradient descent (SGD), proposed in \cite{RobbinsMonro}, has gained a significant popularity due to its simplicity and superior performance. SGD uses a randomly chosen subset of data to obtain a noisy and unbiased estimate of the true gradient. The well-known difficulty with this procedure is that one has to tune its step-size carefully to prevent divergence and a significant amount of work has been devoted to this topic, e.g. see \cite{adaGrad,noMorePesky,mahsereci2015probabilistic}.

A popular alternative to SGD algorithms is the class of incremental proximal methods (IPMs) \cite{bertsekas2011incremental2}. These methods utilize \textit{proximal operators} in an online fashion, i.e., they minimize a single or a mini-batch of components of the cost function in Eq.~\eqref{TheCost} by using a regularizer that depends on the value taken at the previous iteration. Although proximal operators are straightforward to obtain analytically for the linear case, they are not easy to obtain for nonlinear problems. In these cases, every proximal step requires an iterative numerical solver which makes the IPM computationally disadvantegous compared to SGD.

In this paper, we develop a probabilistic interpretation of the IPM for large-scale problems. This approach enables us to obtain purely recursive IPM-type optimizers in the form of approximate filtering algorithms. To attain this goal, we first highlight the relationship between the Kalman filter and the IPM and show that these two algorithms essentially result in very similar update rules for the linear case. Then we discuss how this idea can be extended to nonlinear optimization problems.

Our work is related to an emerging class of algorithms called \textit{probabilistic numerical methods} \cite{hennig2015probabilisticnumerics}. These techniques aim at developing probabilistic models of numerical algorithms, which lead to procedures that explicitly tackle the uncertainties inherent to many numerical problems. Also, there is a body of related work on the usage of stochastic filters for nonlinear optimization. In \cite{bertsekas1996incremental} and \cite{iteratedKalmanAsGauss}, the extended Kalman filter (EKF) is viewed as an incremental Gauss-Newton method. In \cite{stocAppOptFilter}, the author shows that the optimal filter for a linear-Gaussian model can be seen as a stochastic approximation \cite{RobbinsMonro} method. More recently, in \cite{hennig2013quasi}, quasi-Newton algorithms are derived as autoregressive filters. In \cite{probabilisticLMS}, the authors derive the filter for linear regression by obtaining a scalar step-size from the posterior covariance. In \cite{kalman-based-sgd}, the author provides a Kalman-based SGD method which is similar to the algorithm in our linear filtering derivation.
\section{Proximal Operators as Bayes Updates}\label{secProxBayes}
Proximal algorithms have become popular in the signal processing, machine learning, and optimization literature; see  \cite{combettes2011proximal} for a review from a signal processing perspective and  \cite{parikh2014proximal} for a thorough review from an optimization perspective. These algorithms utilize \textit{proximal operators} to move towards the minimum of a cost function. A proximal step can be seen as an implicit gradient step \cite{parikh2014proximal}. Although they are very general, implementing proximal operators is not straightforward and this is seen as the main limitation of these methods.

Consider the proximal operator of a function $f$, defined as
\begin{align*}
\prox_{\lambda,f}(\theta_0) = \argmin_{\theta\in\bR^d} f(\theta) + \lambda g(\theta,\theta_0),
\end{align*}
where $g(\theta,\theta_0)$ is a proper distance and $\lambda \in \bR_+$ is a regularization parameter. This definition leads to the interpretation of proximal methods as majorization-minimization schemes \cite{parikh2014proximal} when $\theta_0$ is replaced by the current estimate $\theta_{k-1}$. Unfortunately proximal operators are not realizable for general $f$ and $g$. Now, if one considers a probabilistic model\footnote{Throughout the paper, $p(x)$ denotes the probability density function (pdf) of a random variable $x$. The notation is argument-wise, e.g., $p(y)$ denotes the pdf of another random variable $y$. The density $p(x|y)$ is the conditional pdf of $x$ given $y$.}
\begin{align*}
p(y | \theta) &\propto \exp(-f(\theta)), \\
p(\theta|\theta_0,\lambda) &\propto \exp(-\lambda g(\theta,\theta_0)),
\end{align*}
where $y$ is the observation implicit in the cost function $f$, we can recover the proximal operator as a maximum-a-posteriori (MAP) estimate, i.e.,
\begin{align*}
\prox_{\lambda,f}(\theta_0)  = \argmax_{\theta\in\bR^d} p(\theta|y,\theta_0,\lambda).
\end{align*}
Although it seems a straightforward observation, this fact brings important implications. Specifically, it means that the family of Bayesian numerical methods can be used to implement proximal operators. Moreover, instead of aiming at the MAP estimate, one can estimate the posterior pdf $p(\theta|y,\theta_0,\lambda)$, which can be used to quantify the uncertainty of the estimate provided by the optimizer \cite{hennig2015probabilisticnumerics}, while the proximal operator has no notion of uncertainty over the solution it provides.

Next, we obtain the proximal operator for the linear regression case explicitly. The following derivation relies on the well-known Bayesian interpretation of the $\ell_2$ regularizer as a Gaussian prior, see e.g. \cite{gribonval2011should}. Nevertheless, we find it useful to state it in the proximal setting.

Consider the proximal operator for a function $f(\theta) = (y - \x^\top \theta)^2$ with the proximal term $g(\theta,\theta_0) = \|\theta - \theta_0\|^2_{2,V^{-1}}$,
\begin{align*}
\tilde{\theta} = \prox_{\lambda,f} (\theta_0) = \argmin_{\theta\in\bR^d} (y - \x^\top \theta)^2 + \lambda \| \theta - {\theta_0}\|_{2,V^{-1}}^2,
\end{align*}
where $y \in \bR, \x \in\bR^d$ and $\theta, {\theta_0} \in \bR^d$ and $V \in \bR^{d\times d}$ is a symmetric positive definite matrix. Then, $\tilde{\theta}$ is given by,
\begin{align}\label{eqProxUpdate}
\tilde{\theta} = \theta_0 + \frac{V \x (y - \x^\top \theta_0)}{\lambda + \x^\top V \x}.
\end{align}
Next, we consider the probability model
\begin{align}
p(y|\theta) &= \NPDF(y; \x^\top \theta, \lambda),\label{eqLik} \\
p(\theta) &= \NPDF(\theta; {\theta_0},V),\label{eqPrior}
\end{align}
where $\NPDF(\z;\mu,S)$ denotes the Gaussian pdf of the vector $\z$ with mean $\mu$ and covariance matrix $S$. Given this model, the posterior pdf can be written as \cite{Bishop2006}
\begin{align}\label{eqPosterior}
p(\theta|y) = \NPDF(\theta; \tilde{\theta},\tilde{V}),
\end{align}
where
\begin{align}
\tilde{\theta} &= \theta_0 + \frac{V \x (y - \x^\top \theta_0)}{\lambda + \x^\top V \x}\label{eqMeanPost}
\end{align}
and
\begin{align}
\tilde{V} &= V - \frac{V \x \x^\top V}{\lambda + \x^\top V \x}.\label{eqCovPost}
\end{align}
One can see that the mean in Eq.~\eqref{eqMeanPost} exactly coincides with the update of Eq.~\eqref{eqProxUpdate}. As a byproduct of the Bayesian update, we get the covariance matrix in Eq.~\eqref{eqCovPost} as a measure of the uncertainty of our estimate.

In the next section, we apply this interpretation to the family of incremental proximal methods in order to get an online and probabilistic optimizer. Perhaps not surprisingly, this algorithm is related to the Kalman filter for the linear case. Then we discuss its extension to nonlinear optimization problems.
\section{Incremental Proximal Methods}
IPMs are a class of algorithms that aim at solving problems of the form of Eq.~\eqref{TheCost} by using only a single component $f_k$ at each iteration \cite{bertsekas2011incremental2}. Given the estimate of the minimum $\theta_{k-1}$, the IPM generates the next estimate by solving the following problem:
\begin{align}
\theta_k &= \prox_{\lambda,f_k} (\theta_{k-1}) = \argmin_{\theta\in\bR^{d}} f_k(\theta) + \lambda \|\theta - \theta_{k-1}\|_{2,V^{-1}}^2, \label{ExplicitProxOp}
\end{align}
which uses only the most recent ``observation''. Let us abuse the notation a bit and let $f_k$ stand for $f_{i_k}$ where $i_k$ is sampled uniformly randomly from the index set $[n] = \{1,\ldots,n\}$ (the same as for SGD algorithms). The matrix $V$ is usually selected as the identity matrix but other choices, depending on the geometry of the problem, are possible. When the problem in Eq.~\eqref{ExplicitProxOp} is solvable, it is argued that it leads to more stable algorithms than the SGD (see \cite{bertsekas2011incremental2} for a discussion on convergence).
\subsection{The IPM as a Kalman filter for linear regression}
Given an observation vector $\y \in \bR^n$ and a feature matrix $X \in \bR^{d\times n}$, the linear regression problem consists in fitting a vector $\theta \in \bR^{d}$ which roughly satisfies $\y \approx X^\top \theta$. Formally, the problem can be framed as the minimization of $f(\theta) = \|\y - X^\top\theta\|_2^2$. When $n$ is large, solving this problem analytically becomes unfeasible. Hence, stochastic optimization methods are often applied. We note that, in this setting, $f(\theta) = \sum_{k=1}^n f_k(\theta)$ and each component $f_k$ can be written as $f_k(\theta) = (y_k -\x_k^\top \theta)^2$ where $y_k \in \bR$ is a single observation and $\x_k$ is a column of the feature matrix $X$. Then, as shown in Section~\ref{secProxBayes}, the incremental proximal iteration can be written as
\begin{align}\label{IPMUpdate}
\theta_k = \theta_{k-1} + \frac{V \x_k (y_k - \x_k^\top \theta_{k-1})}{\lambda + \x_k^\top V \x_k}.
\end{align}
For a proper Bayesian interpretation, we would like to obtain Eq.~\eqref{IPMUpdate} as a recursive posterior-mean update in a (Gaussian) probabilistic model. In this case, the probabilistic updates yield to a similar, but \textit{different}, algorithm. Notice that, in the probabilistic interpretation of Section~\ref{secProxBayes}, $V$ denotes the posterior covariance. However, in Eq.~\eqref{IPMUpdate}, this matrix is kept constant through iterations, as there is no way to update it. In the online optimization literature, some algorithms are proposed to update this matrix, which amounts to updating the proximal term, such as AdaGrad \cite{adaGrad}. As we show below, in the probabilistic approach, the matrix $V$ is naturally updated as the posterior covariance matrix.

Let us consider the model
\begin{align*}
p(\theta) = \NPDF(\theta; \theta_0, V_0), \,\,\,\,\,\,\,\,\,\, p(y_k | \theta) &= \NPDF(y_k; \x_k^\top \theta, \lambda).
\end{align*}
Given the data sequence $y_{1},\ldots,y_k$, the posterior distribution $p(\theta | y_{1:k})$ is Gaussian \cite{Bishop2006}. We denote it as $p(\theta|y_{1:k}) = \NPDF(\theta; \theta_k, V_k)$. The mean $\theta_k$ and covariance $V_k$ can be computed as \cite{anderson1979optimal}
\begin{align}
\theta_k &= \theta_{k-1} + \frac{V_{k-1} \x_k (y_k - \x_k^\top \theta_{k-1})}{\lambda + \x_k^\top V_{k-1} \x_k}, \label{probIPPmeanUpdate}\\
V_k &= V_{k-1} - \frac{V_{k-1} \x_k \x_k^\top V_{k-1}}{\lambda + \x_k^\top V_{k-1} \x_k}. \label{probIPPCovUpdate}
\end{align}
The relationship between the Eqs.~\eqref{IPMUpdate} and \eqref{probIPPmeanUpdate} is evident. In the probabilistic approach, we have $V_k$ instead of $V$, which means that we obtain a way to \textit{update the proximal term} in the Eq.~\eqref{ExplicitProxOp} in a principled way and with an intuitive meaning ($V_k$ quantifies the uncertainty in the solution $\theta_k$).
%
\subsection{Extended Kalman filter as an IPM for nonlinear regression}\label{secEKF}
In this section, we consider a nonlinear regression problem. Given observations $\y$, we would like to obtain $y_k \approx g(x_k,\theta)$ where $g(\cdot,\theta)$ is a nonlinear function of $\theta$. Since the $x_k$'s are fixed, we set $g_k(\theta) := g(x_k,\theta)$ for notational conciseness. Note that $g_k:\bR^{d} \to \bR$. Then, we would like to solve the problem
\begin{align}\label{nonlinearReg}
\min_{\theta\in\bR^d} f(\theta) = \min_{\theta\in\bR^d} \sum_{k=1}^n (y_k - g_k(\theta))^2.
\end{align}
The incremental proximal step for this problem is
\begin{align}\label{eqProxStepNonLinear}
\theta_k= \argmin_{\theta\in\bR^d} (y_k - g_k(\theta))^2 + \lambda \|\theta - \theta_{k-1}\|_{2,V^{-1}}^2
\end{align}
for each iteration $k$. Because this proximal step is intractable in general, the typical choice for problems like in Eq.~\eqref{nonlinearReg} is the SGD. In what follows, we propose the use of EKF recursions as one-step approximations of the realization of the proximal operator.

To this end, let us consider the probabilistic model
\begin{align}\label{nonlinearGaussModel}
p(\theta) = \NPDF(\theta;\theta_0,V_0),\,\,\,\,\,\,\,\,\,\, p(y_k|\theta) &= \NPDF(y_k; g_k(\theta),\lambda).
\end{align}
Since the model is nonlinear, using the EKF is a natural way to solve the regression problem. Let us denote $h_k = \nabla_\theta g_k(\theta_{k-1})$. Then, the EKF recursions can be written as,
\begin{align}\label{ExtendedKalmanIPP}
\theta_k = \theta_{k-1} + \frac{V_{k-1} h_k (y_k - g_k(\theta_{k-1}))}{\lambda + h_k^\top V_{k-1} h_k}
\end{align}
and
\begin{align*}
V_k = V_{k-1} - \frac{V_{k-1}h_k h_k^\top V_{k-1}}{\lambda + h_k^\top V_{k-1} h_k}.
\end{align*}
Note that these updates are different from the ones that would be obtained from a naïve linearization of $g_k$ followed by the derivation of the IPM. For that case, we would have the term  $(y_k - h_k^\top \theta_{k-1})$, instead of $(y_k - g_k(\theta_{k-1}))$, which does not ensure numerically stable updates.

\subsection{Nonstationary optimization}
Throughout our discussion, we have kept the prior $p(\theta)$ static, meaning that $\theta$ is assumed to be random but not changing with time. While this assumption is convenient when the cost function is not changing, it does not hold in most realistic settings. For this reason, the authors of \cite{noMorePesky} consider what they call \textit{nonstationary} losses, where the cost function is also changing with time. Tackling such a scenario  is trivial from our perspective, as one only needs to modify the algorithm slightly in order to get a dynamic algorithm. In particular, in addition to the update step, one needs to employ a prediction step, according to the assumed dynamics of the parameter. One can model the degree of nonstationarity by modifying the model over $\theta_k$ and filtering algorithms extend to such settings very naturally. We leave the detailed investigation of this aspect for future work.
\section{Numerical Results}
In this section, we investigate two algorithms on a simple problem of fitting a sigmoid function. The first algorithm, which we refer to as \textit{approximate nonlinear IPM}, consists of applying a standard iterative solver for each subiteration since the nonlinear problem of Eq.~\eqref{eqProxStepNonLinear} is not solvable in general. The second algorithm is the EKF, as explained in Section~\ref{secEKF}. The model used in the experiment is of form Eq.~\eqref{nonlinearGaussModel} with,
\begin{align*}
g_k(\theta) = \frac{1}{1 + \exp(-\alpha - \beta^\top x_k)}
\end{align*}
where $x_k \in \bR^{d-1}$ denotes the inputs, and the parameter vector is $\theta = (\alpha,\beta)$ where $\theta \in \bR^{d}$ with $d = 21$. Recall that, by choosing such a model, we aim at solving a problem of form given in Eq.~\eqref{nonlinearReg}. We set the value of the parameter $\lambda = 0.2$ while generating the data and we use the same value in algorithms. Also, the initial value $\theta_0$ of the approximate IPM is set randomly while the proximal matrix is the $d\times d$ identity, denoted $V = I_d$. Similarly, the prior for the EKF is initialized with $(\theta_0,V_0)$ where $V_0 = I_d$.
\begin{figure}[t]
\begin{center}
\includegraphics[scale=0.45]{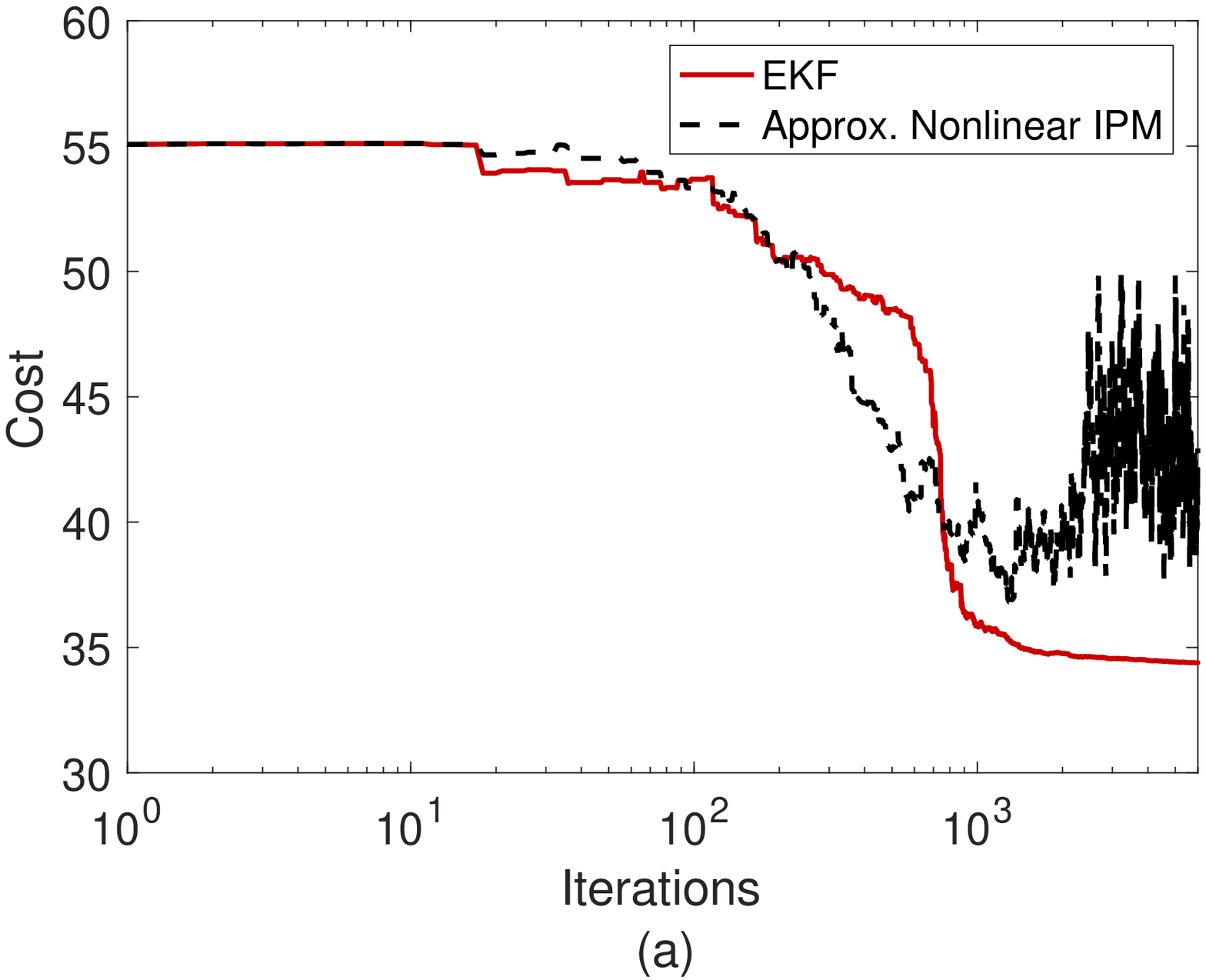}
\includegraphics[scale=0.22]{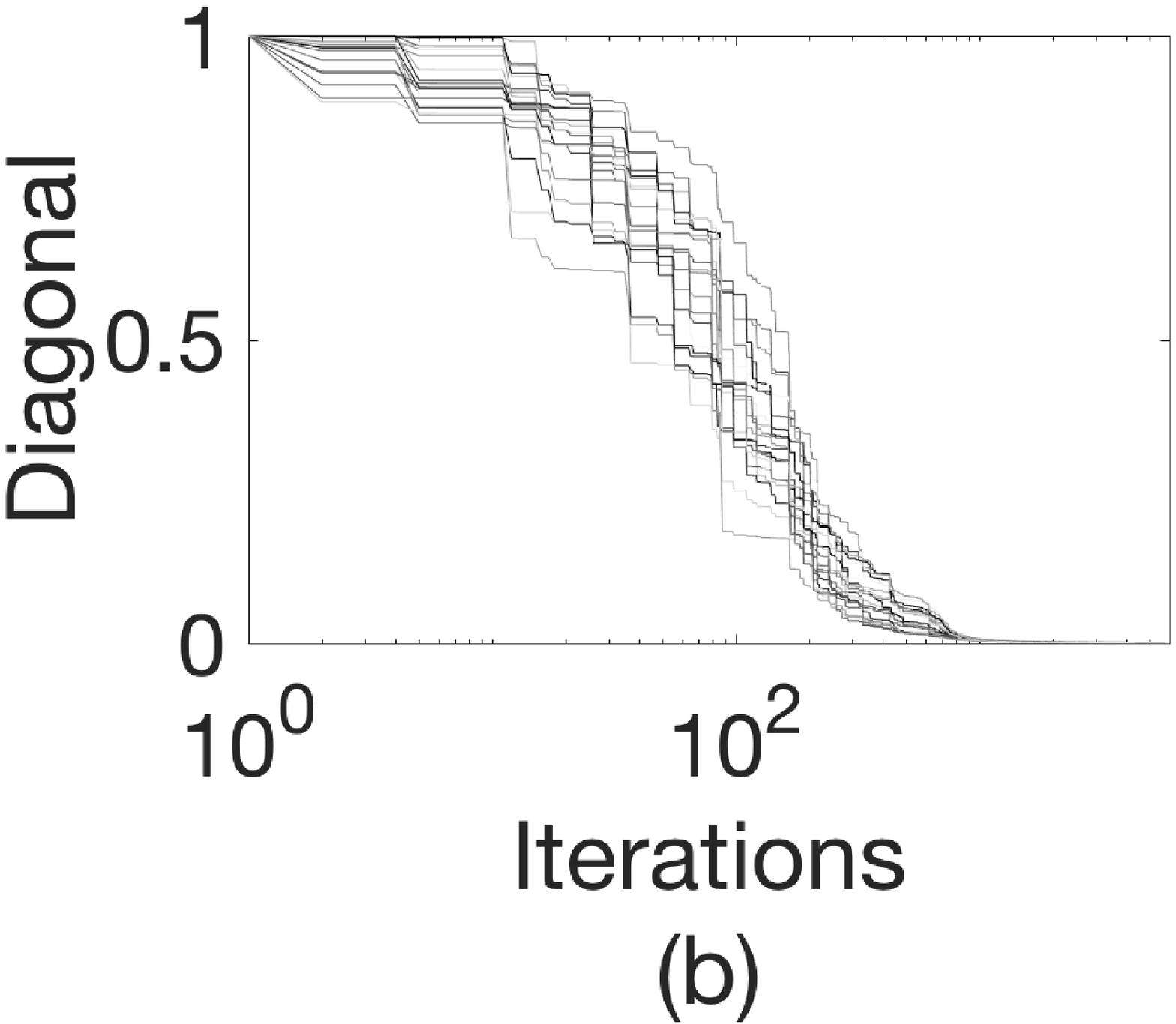}
\includegraphics[scale=0.22]{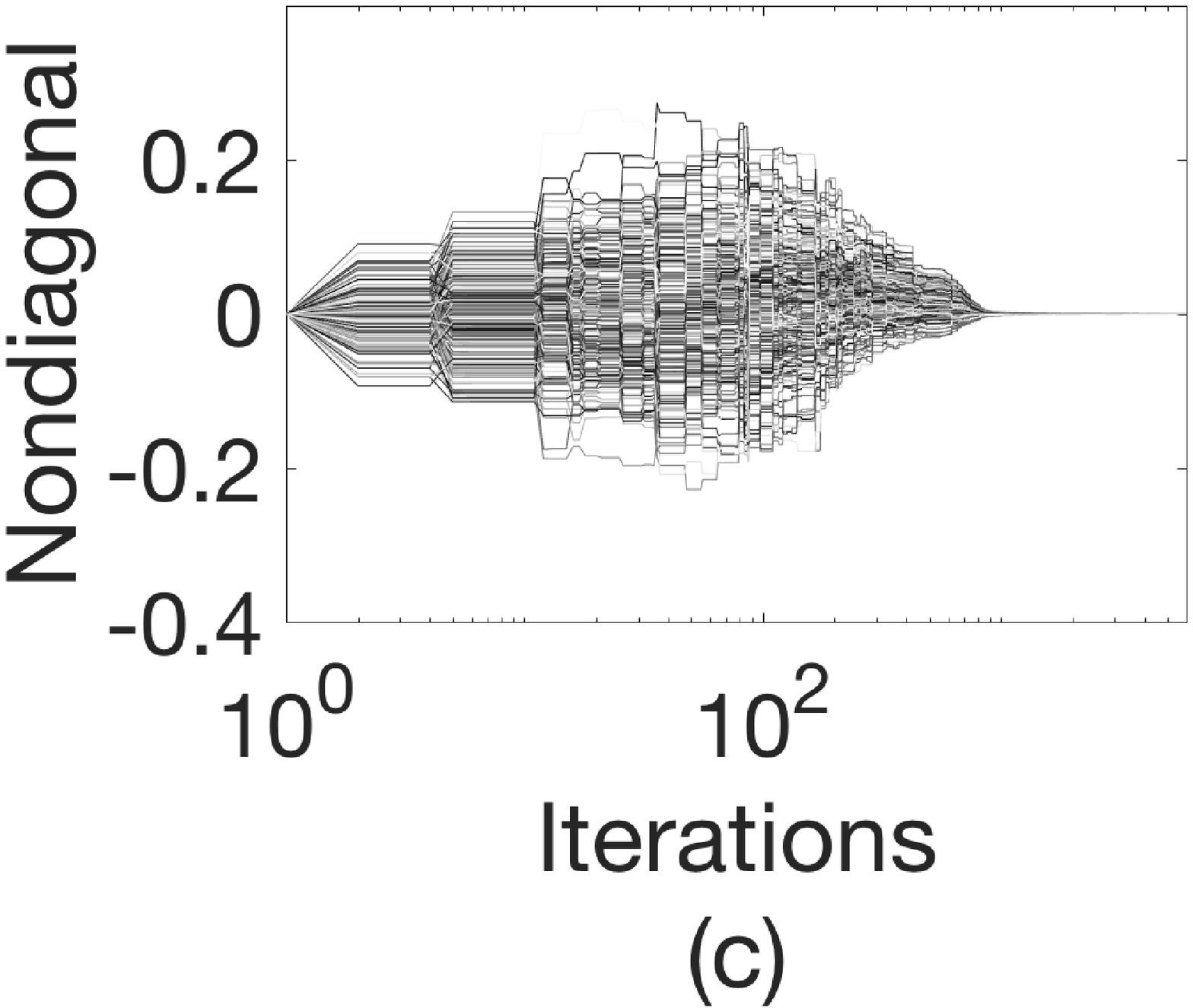}
\end{center}
\caption{Results on fitting a sigmoid function using EKF and approximate nonlinear IPM. From (a), it can be seen that the approximate nonlinear IPM proceed towards minimum but suffers from instability while the EKF proceeds in a stable way. From (b)-(c), it can be seen that the entries of the diagonal and nondiagonal parts of the covariance matrix $V_k$ converge to zero which is the reason why the EKF does not suffer from instability.}
\label{FigComp}
\end{figure}

Figure~\ref{FigComp}(a) shows that the approximate IPM suffers from numerical instability as the parameter estimate $\theta_k$ becomes close to the actual minimum. One reason for this instability is that, in the proximal-type algorithms, there is no natural mechanism to reduce the size of the step taken by the algorithm. A natural remedy would be to update the proximal matrix in a way that it dampens the updates as the number of iterations increases, in a similar way to decreasing the step-size of the SGD. The EKF exactly employs this strategy in a natural way. This fact can be seen from Fig.~\ref{FigComp}(b)-(c) where the diagonal and nondiagonal entries of the covariance matrix $V_k$ are plotted, respectively. It is evident that the entries of this matrix converge to zero, meaning that the update \eqref{ExtendedKalmanIPP} eventually converges to some point in the parameter space.

\section{Conclusions}
In this work, we have developed a probabilistic perspective for proximal and incremental proximal methods. We have shown that a probabilistic setting can provide a systematic way to derive algorithms when this is not possible from the classical perspective. In particular, within an online setup, we have argued that the use of filtering algorithms corresponds to employing an IPM-type scheme for optimization. However, filtering algorithms have natural dampening mechanisms for parameter updates, as they refine their uncertainty over the solution iteratively.

This line of work can be pushed forward in a number of different directions. First, different Kalman filters can be used in a similar way to get more advanced optimization schemes for more complicated problems. Among the candidates, the unscented Kalman filter (UKF) and the Ensemble Kalman filter (EnKF) \cite{sarkka2013bayesian} can be useful to tackle high dimensional, possibly time-varying, optimization problems. Second, this approach can be extended beyond quadratic functions by exploiting the relationship between exponential families and Bregman divergences \cite{clusteringwithBreg}. When the likelihood belongs to the exponential family, the cost function can be expressed in general as a Bregman divergence. In that case, since the Gaussianity assumption is violated, one needs to resort to more complicated numerical algorithms, such as particle filters \cite{djuric2003particle} or other advanced filtering methods.

\subsubsection*{Acknowledgements}
{\"O}.~D.~A. and J.~M. acknowledge the support of \textit{Ministerio de Econom\'ia y Competitividad} of Spain (TEC2015-69868-C2-1-R ADVENTURE), the Office of Naval Research Global (N62909-15-1-2011), and the regional government of Madrid (program CASICAM-CM S2013/ICE-2845). V.~E. acknowledges support from the \textit{Agence Nationale de la Recherche} of France under PISCES project (ANR-17-CE40-0031-01).
\bibliographystyle{IEEEbib}
\bibliography{draft}

\end{document}